\documentclass[floats,preprint,prb,aps]{revtex4}
\usepackage{graphicx}
\usepackage{epsfig}

\begin{document}
\preprint{\emph{Submitted to Phys. Rev. \textbf{A}}}

\title{Photon Statistics; Nonlinear Spectroscopy of Single Quantum Systems}

\author{Shaul Mukamel}
\affiliation{Department of Chemistry, University of California,
Irvine, CA 92697-2025}

\date{\today}

%-------------------------------------------------------------------------
%  ABSTRACT
%-------------------------------------------------------------------------
\begin{abstract}
A unified description of multitime correlation functions,
nonlinear response functions, and quantum measurements is
developed using a common generating function which allows a direct
comparison of their information content. A general formal
expression for photon counting statistics from single quantum
objects is derived in terms of Liouville space correlation
functions of the material system by making a single assumption
that spontaneous emission is described by a master equation.

\vspace{1.0 in} Pacs numbers: 42.50.Ar, 42.50.Lc, 42.65.-k,
03.65.Ta

\end{abstract}

\maketitle

%\tableofcontents
%-------------------------------------------------------------------------
%  MAIN TEXT
%-------------------------------------------------------------------------
\newpage
\section{Introduction}

Knowledge of the full density matrix of the radiation field allows
to compute all its measurable properties. In particular, photon
counting statistics which had proven to be a most valuable measure
of coherence has been formulated in the sixties by Kelley and
Kleiner, Glauber, and Mandel in terms of expectation values of
normally-ordered field
operators~\cite{Kelley,Glauber,Glauber2,Glauber3,Mandel,MandelBook,Mandel2,Ackerhalt}.
A revived interest had emerged in the eighties when stochastic
trajectory experiments on single quantum objects, atoms, ion
traps, molecules and quantum dots became
feasible~\cite{Leibfried,Cook,Diedrich,Kimble,Schubert,Basche,moerner,VandenBout,Jung}.
It is more convenient to formulate such measurements in terms of
correlation functions of the material system, rather than of the
radiation field. Various applications for specific few-level model
systems have been
investigated~\cite{Leibfried,Carmichael,Blatt,Metz,Zumofen}.

In this paper we develop correlation function expressions for
photon statistics which apply to a general model of a quantum
system driven by an external field and coupled to a bath. The
normally-ordered field expressions are remarkably general; the
only assumption made in their derivation is that the photon
detection is described by the Fermi golden rule. Similarly, our
material correlation function expressions hold under a single
assumption; that spontaneous emission can be described by a master
equation. The derivation of the master equation starting with the
fully quantum description of the field is well
documented~\cite{43,AgarwalBook,Lehmberg}. No other properties of
the radiation field enter explicitly in the present formulation.
Computing the reduced density matrix of a single quantum system
coupled to a bath has been a long standing goal of nonequilibrium
statistical mechanics~\cite{43,Zwanzig,Kubo}. Various types of
reduced equations of motion based on stochastic or microscopic
models are well developed. The present approach is therefore
particularly useful for single quantum systems since it can
utilize any level of reduced equations of motion to predict the
photon statistics. Computing the many-body density matrix of a
macroscopic system is much more complex and a collective
description using field
operators~\cite{Kelley,Glauber,Glauber2,Glauber3,Mandel,MandelBook,Mandel2,Ackerhalt}
may then be more adequate.

We further develop some fundamental connections between multitime
correlation functions of photon statistics and response functions
of nonlinear spectroscopy~\cite{shaul}. Coherent experiments
conducted using multiple pulses provide a wealth of information on
electronic and nuclear dynamics~\cite{42b}. These techniques can
create and manipulate quantum coherences among selected states and
the signals provide snapshots of their dynamics.  There are
numerous motivations for performing such measurements. (i) Novel
spectroscopy: the ability to explore and access unusual regions of
phase space. (ii) Coherent control: achieving a desired goal (e.g.
optimize the branching ratio of a reaction towards a favorable
product).  (iii) Quantum computing: these applications depend on
generating and retrieving information about coherences between
several degrees of freedom prepared in correlated wavepackets
(such {\em entanglement} is a synonym to the old fashioned term
{\em correlation}).  (iv) Overcome coupling to a bath e.g.
selectively eliminating {\em dephasing} processes(or the more
trendy term {\em decoherence}). We shall draw upon the analogy
between photon statistics and nonlinear response and correlation
functions to show important similarities and differences in their
information content and simulation strategies.

In Section II we present the Liouville space expressions for
multipoint correlation functions and response functions. In
Section III we discuss multipoint quantum equilibrium measurements
and introduce a unified generating function that can be used to
compute correlation, response and measurements. This sets the
stage for deriving the Liouville space expressions for photon
counting in Section IV. Finally, our results are summarized in
Section V.

\section{Multipoint Generating Functions for Correlation and Response Functions}

The state of complex quantum systems may be conveniently
characterized by multitime quantities which carry various levels
of information and are easier to calculate, measure, or visualize
compared to the many-body wavefunction~\cite{b}. In this section
we present formal expressions for two such objects (correlation
and response functions) using a Liouville space (superoperator)
approach. While the results of this section are not new, they
establish the notation, setting the stage for calculating the
successive measurements in Section III and eventually to the
photon statistics in Section IV, which are our main results.
Correlation and response functions can be defined in Hilbert space
using ordinary operators. Quantum measurements on the other hand,
must be formulated using density matrices in Liouville space. The
notation introduced here is essential for expressing all of these
quantities using a common generating function (Eq.~(\ref{25}))
which unambiguously reveals their relative information content.

Consider a dynamical variable of interest $A$ which can represent
e.g. the dipole operator, the coordinate or the momentum of a
tagged particle, or some collective coordinate. The simplest
$n$-point object is the correlation function
%----1------------------------------------------------------
\begin{equation}\label{simplest}
C^{(n)} (\tau_{n}\ldots \tau_{n})\equiv
\mathrm{Tr}[A(\tau_{n})\ldots A(\tau_{1})\rho_{eq}],
\end{equation}
%-----------------------------------------------------------
where $\rho_{eq}$ is the equilibrium density matrix of the system.

Classical correlation functions are given by moments of the joint
distribution of successive measurements and are therefore directly
observable. Quantum correlation functions, in contrast, are not
connected to specific measurements in a simple way. Instead,
response functions~\cite{shaul,b,Ernst,Zwanzig} which represent
the reaction of the system to an external field $E(t)$ coupled to
the variable $A$ via $H_{int} = -E(t)\cdot A$ may be readily
measured. In a response experiment the total Hamiltonian
$H_T(\tau)$ consists of material Hamiltonian $H$ and the coupling
to the driving field
%----2-----------------------------------------------------------------
\begin{equation}\label{2.4t}
H_T(\tau) = H + H_{int}(\tau).
\end{equation}
%----------------------------------------------------------------------
We shall be interested in the expectation value of $A$ at time $t$

%-----3-------------------------------------------------------
\begin{equation}
\label{expectation} U (t) = \mathop{\rm Tr} [ A \rho (t)]
 \equiv \langle\langle  A \big| \rho (t) \rangle\rangle,
\end{equation}
%-------------------------------------------------------------
where $\big| \rho (t) \rangle\rangle = \sum_{jk} \rho_{ik}(t)
\big| jk \rangle\rangle$ is the density matrix of the system, and
the ket $|jk\rangle\rangle$ denotes the Liouville-space operator
$|j \rangle \langle k |$ ~\cite{shaul,Fano,Zwanzig}.

Eq.~(\ref{expectation}) can be recast in the
form~\cite{Mukamel,Wang}
%------4---------------------------------------------------------------
\begin{equation}\label{3.3t}
  U(t) = \langle T A_+(t) \exp \left [\frac {i} {\hbar} \int_{-\infty}^{t} d\tau
  E(\tau)A_{-}(\tau) \right] \rangle.
\end{equation}
%----------------------------------------------------------------------
Here and below $\langle \cdots \rangle$ denotes averaging with
respect to the equilibrium density matrix $\rho_{eq}$
%-------5-------------------------------------------------------------
\begin{equation}
\langle A \rangle\equiv \mathrm{Tr}[A\rho_{eq}],
\end{equation}
%----------------------------------------------------------------------
$A_\pm$ are superoperators acting in Liouville space defined as
follows: For any ordinary operator $A$ we define
%-------6--------------------------------------------------------------
\begin{equation}\label{A3t}
A_- \equiv A_L - A_R; \hspace{0.2in} A_+ \equiv \frac {1} {2} (A_L
+ A_R),
\end{equation}
%----------------------------------------------------------------------
where $ A_L$ and $ A_R$ are the superoperators that act on the ket
(left) and bra (right) of the density matrix ($A_L B \equiv AB$
and $A_R B \equiv -BA_R$).

The time evolution in Eq.~(\ref{3.3t}) is given in the interaction
picture. For an ordinary operator in Hilbert space this is defined
by
%--------7-------------------------------------------------------------
\begin{equation}
\label{copied6a} A(\tau) \equiv
\exp\left(\frac{i}{\hbar}H\tau\right)A \exp
\left(-\frac{i}{\hbar}H\tau\right).
\end{equation}
%----------------------------------------------------------------
Similarly, the time evolution of superoperators is governed by the
Liouville operator  $H_{-}$ corresponding to the material
Hamiltonian $H$
%---------8------------------------------------------------------------
\begin{equation}
\label{6a} A_{j}(\tau) \equiv
\exp\left(\frac{i}{\hbar}H_{-}\tau\right)A_{j} \exp
\left(-\frac{i}{\hbar}H_{-}\tau\right). \hspace{0.2in}j=+,-,L,R
\end{equation}
%----------------------------------------------------------------
$T$ is the positive time ordering operator which rearranges all
products of superoperators in order of decreasing time from left
to right. The nonlinear response functions are obtained by
expanding the exponent of Eq.~(\ref{3.3t}) in powers of $E(\tau)$.
The expectation value of $A$ to $n-1$'th order in the field is
given by
%---------9------------------------------------------------------------
\begin{equation}\label{3.4t}
U^{(n-1)}(\tau_{n}) = \int_{-\infty}^{\tau_{n}}
dt_{\tau_{n-1}}\ldots
 \int_{-\infty} ^{\tau_2} d{\tau_1} R ^{(n)}
 (\tau_{n}\ldots \tau_1) \nonumber \\
   {\cal E} (\tau_{n-1})\ldots {\cal E}
   (\tau_1),\hspace{0.2in}n=2,3\ldots
\end{equation}
%----------------------------------------------------------------------
with the \emph{nonlinear response function}
%----------10-----------------------------------------------------------
\begin{equation}\label{3.5t}
  R^{(n)} (\tau_{n}\ldots \tau_1) \equiv
  \left ( \frac {i} {\hbar} \right )^n
  \langle A_+ (\tau_{n}) A_- (\tau_{n-1}) \ldots A_-(\tau_1) \rangle.
\end{equation}
%----------------------------------------------------------------------
$R^{(n)}$ represents the response at times $\tau_{n}$ to $n-1$
very short pulses applied at times
$\tau_{1}\cdots\tau_{n-1}$~\cite{shaul}. Note that all time
arguments are fully ordered $\tau_1 \leq \tau_2 \ldots \leq
\tau_{n}$. The operator $A_+(\tau_{n})$ corresponds to the
observation time, the operators $A_{-}(\tau_j)$ $j=1,\cdots, n-1$
represents interactions with the external field at times
$\tau_{j}$. We chose to label the response function corresponding
to $U^{(n-1)}$ by $R^{(n)}$ rather than $R^{(n-1)}$ since it is an
$n$ point function; this will facilitate the comparison with the
other multitime quantities discussed below.

Eq.~(\ref{3.5t}) is an abbreviated notation for
%------11+12--------------------------------------------------------------
\begin{equation}\label{3.6t}
R^{(n)} (\tau_{n} \ldots \tau_1) = \left( \frac {i}{\hbar}
\right)^n
\nonumber \\
\langle[\ldots[[A(\tau_{n}), A(\tau_{n-1})], A(\tau_{n-2})]\ldots,
A(\tau_1)] \rangle,
\end{equation}
%----------------------------------------------------------------------
or
%--------13------------------------------------------------------------
\begin{equation}
\label{eq2}
R^{(n)} (\tau_{n} \ldots \tau_1)  = \!\bigg( \!\frac {i}{\hbar}
\!\bigg)^{\!n} {\rm Tr} \Big\{ A(\tau_{n}) \Big[ A(\tau_{n-1}),
\dots, [A(\tau_2), [ A(\tau_1), \rho_{eq}]] \cdots \Big] \Big\}.
\end{equation}
%-------------------------------------------------------------

$R^{(n)}$ is thus given by a combination of $n$-order ordinary
(Hilbert space) correlation functions. Eq.~(\ref{eq2}) contains
$2^{n-1}$ terms representing all possible ``left'' and ``right''
actions of the various commutators. Each term corresponds to a
\emph{Liouville-space path} and can be represented by a
double-sided Feynman diagram~\cite{shaul}. The various pathways
interfere, giving rise to many interesting effects such as new
resonances. For a multilevel system $R^{(n)}$ is usually expanded
in the eigenstates of the free Hamiltonian $H$. Each path then
consists of $n-1$ periods of free evolution separated by $n$
couplings with $A$, which change the state of the system.

Using our superoperator notation, the correlation function
Eq.~(\ref{simplest}) can be recast as
%-------14---------------------------------------------------
\begin{equation}
\label{12a} C^{(n)} (\tau_{n}\ldots \tau_{1})=
\mathrm{Tr}[A_{L}(\tau_{n})\ldots A_{L}(\tau_{1})\rho_{eq}],
\end{equation}
%-----------------------------------------------------------
where the time evolution of $A_{L}(\tau)$ is given by
Eq.~(\ref{6a}).

Classical quantities are conveniently represented as moments of
some \emph{joint distribution functions} connected to
measurements. The closest we can come up in quantum mechanics is
through moments of \emph{generating functions}. This is not only a
convenient computational tool, but also provides insights and
helps connect different measurements. The generating function for
correlation functions is defined as
%------15------------------------------------------------------
\begin{equation}\label{Eq. 1}
W_{C}^{(n)} (a_{n}\tau_{n}\cdots a_{1}\tau_{1})\equiv\langle
\delta (a_{n}-A(\tau_{n}))\cdots\delta (a_{1}-A(\tau_{1}))\rangle.
\end{equation}
%-------------------------------------------------------------
By comparing Eq.~(\ref{simplest}) and Eq.~(\ref{Eq. 1}) we
immediately find that
%------16----------------------------------------------------
\begin{equation}\label{simplest2}
C^{(n)} (\tau_{n}\ldots \tau_{1})= \int \ldots \int a_{1}\ldots
a_{n} W_{C}^{(n)}(a_{n}\tau_{n}\cdots
a_{1}\tau_{1})\,\,da_{1}\ldots da_{n}.
\end{equation}
%-----------------------------------------------------------
Similarly we can define a generating function for response
functions
%----17------------------------------------------------------
\begin{eqnarray}
\label{14} & &W_{R}^{(n)}(a_{n}\tau_{n}\cdots
a_{1}\tau_{1})\\\nonumber & &=
\bigg\langle[\delta(a_{n}-A(\tau_{n})),\cdots[\delta(a_{2}-A(\tau_{2})),
[\delta(a_{1}-A(\tau_{1})),\rho_{eq}]]]\bigg\rangle,
\end{eqnarray}
%-----------------------------------------------------------
so that the response function is given by
%--------18-------------------------------------------------
\begin{eqnarray}
\label{16} R^{(n)}(\tau_{n} \ldots \tau_1) = \int \ldots \int
a_{1}\cdots a_{n} W_{R}^{(n)}(a_{n}\tau_{n}\cdots
a_{1}\tau_{1})\,\,da_{1}\cdots da_{n}.
\end{eqnarray}
%-----------------------------------------------------------

Eqs.~(\ref{Eq. 1}) and~(\ref{14}) play the role of a classical
distribution functions even though they are generally complex and
may be negative. Nevertheless, they serve as generating functions
for correlation and response functions which are given by their
first ``moments'', Eqs.~(\ref{simplest2}) and~(\ref{16}).

So far we considered two $n$ point objects: Correlation functions
and response functions. A third type of quantity which is more
closely connected to photon statistics is the joint distribution
of $n$ successive measurements. This will be introduced next.

\section{Unified Generating Function for Correlation, Response and
Equilibrium Measurements}

We consider a sequence of $n$ measurements of a dynamical variable
$A$ performed on the system at times
$\tau_{1}\leq\tau_{2}\leq\tau_{3}\cdots\leq\tau_{n}$ and yielding
the outcomes $a_{1}\cdots a_{n}$. We would like to compute the
following ensemble average over many such measurements
%--------------------------------------------------------------
\begin{eqnarray}\label{ensemble}
S^{(n)}\tau_{n}\cdots\tau_{1}\equiv \overline{A(\tau_{n})\cdots
A(\tau_{1})}.
\end{eqnarray}
%-----------------------------------------------------------
This quantity is common to classical and quantum systems alike.
Eq~(\ref{ensemble}) is a shorthand notation for
%--------------------------------------------------------------
\begin{eqnarray}
& &S^{(n)}(\tau_{n}\cdots\tau_{1})=\\\nonumber & &\int \cdots \int
a_{1}\cdots a_{n}\,\,W_{S}^{(n)}(a_{n}\tau_{n},\cdots,
a_{1}\tau_{1})\,\,da_{1}\cdots da_{n}
\end{eqnarray}
%-----------------------------------------------------------
where the joint distribution function $W_{S}^{(n)}$ can be
computed using the theory of quantum
measurements~\cite{Neumann,gellmann,griffiths,omnes1,omnes2}. To
that end, we define the eigenstates $|\alpha_{j}>$ of $A$ with
eigenvalues $a_{j}$, and represent the operator $A$ in the form of
an expansion in projection operators $\hat{A}_{j}$
%-------------------------------------------------------
\begin{equation}
A = \sum_{j}a_{j}\hat{A}_{j},
\end{equation}
%-------------------------------------------------------
with
%-------------------------------------------------------
\begin{equation}
\hat{A}_{j}\equiv|\alpha_{j}><\alpha_{j}|.
\end{equation}
%-------------------------------------------------------

We next define the Liouville space projection operator onto the
diagonal elements of $A$
%-----------------------------------------------------------
\begin{equation}
\label{18} \hat{P}({a}) \equiv
\sum_{j}\delta(a-a_{j})|\hat{A}_{j}\gg \ll \hat{A}_{j}|.
\end{equation}
%-----------------------------------------------------------
The Liouville space bracket $\ll F|G \gg\equiv\mathrm{Tr}'FG$
denotes a scalar product computed by a partial trace over the
measured degrees of freedom of the operator $A$. Using this
projection, the joint distribution function of successive
measurements may be recast in the
form~\cite{griffiths,omnes1,omnes2,Schultz,Neumann}

%-------------------------------------------------------
\begin{equation}
\label{19} W_{S}^{(n)}(a_{n}\tau_{n}\cdots a_{1}\tau_{1})=
\mathrm{Tr} \left[\hat{P}(a_{n}, \tau_{n})\cdots
\hat{P}(a_{1},\tau_{1})\rho_{eq}\right]
\end{equation}
%-------------------------------------------------------
where the time evolution of $\hat{P}(a,\tau)$ is given by the
interaction picture (Eq.~(\ref{6a})). The compact Liouville space
notation used in Eq.~(\ref{19}) will help establish the connection
between photon counting and other multitime quantities. Note that
both $W_{C}$ and $W_{S}$ are normalized as
%-------------------------------------------------------
\begin{equation}
\int W_{j}^{(n)}(a_{n}\tau_{n}\cdots a_{1}\tau_{1})\,\,
da_{1}\cdots da_{n}=1 \hspace{0.2in} j=C,S
\end{equation}
%----------------------------------------------------------
whereas $W_{R}$, which represents the deviation of the density
matrix from equilibrium, has a zero trace~\cite{Mukamel2}
%-------------------------------------------------------
\begin{equation}
\int W_{R}^{(n)}(a_{n}\tau_{n}\cdots a_{1}\tau_{1})\,\,
da_{1}\cdots da_{n}=0.
\end{equation}
%----------------------------------------------------------

So far, we introduced three different generating functions to
describe correlation functions, response functions and the joint
probability of measurements (Eqs.~(\ref{Eq. 1}),(\ref{14}) and
(\ref{19}), respectively). To establish the connection between
these quantities it will be useful to compute all three using a
single fundamental object. This is possible by using the following
generating function,
%-------------------------------------------------------------
\begin{eqnarray}
\label{25} & W^{(n)} (a_{n} a_{n}^{\prime} \tau_{n},a_{2}
a_{2}^{\prime} \tau_{2}\cdots a_{1} a_{1}^{\prime} \tau_{1})
\equiv\nonumber\\
& \mathrm{Tr} \left\{\delta (a_{n}-A (\tau_{n})) \cdots \delta
(a_{1}-A (\tau_{1})) \rho_{eq} \delta (a_{1}^{\prime}-A
(\tau_{1}))\cdots \delta (a_{n}^{\prime}-A (\tau_{n}))\right\}.
\end{eqnarray}
%-------------------------------------------------------------
The density matrix underlying Eq.~(\ref{25}) is
%-------------------------------------------------------------
\begin{equation}
\label{26} \rho^{(n)}(a_{n} a_{n}^{\prime} \tau_{n}\cdots a_{1}
a_{1}^{\prime} \tau_{1})= \mathcal{G}_{a_{n}a_{n}^{\prime} a_{n-1}
a_{n-1}^{\prime}} (\tau_{n}-\tau_{n-1}) \cdots
\mathcal{G}_{a_{2}a_{2}^{\prime}a_{1} a_{1}^{\prime}}
(\tau_{2}-\tau_{1})\rho_{a_{1}a_{1}^{\prime}} (\tau_{1}),
\end{equation}
%-------------------------------------------------------------
where
%-------------------------------------------------------------
\begin{equation}
\label{100a} \mathcal{G}(\tau) \equiv \exp
\left(-\frac{i}{\hbar}H_{-}\tau\right),
\end{equation}
%-------------------------------------------------------------
is the interaction-picture propagator.

The generating function for correlation functions (Eq.~(\ref{Eq.
1})) is recovered by integrating Eq.~(\ref{25}) over the primed
variables
%-------------------------------------------------------------
\begin{equation}
W_{C}^{(n)} (a_{n}\tau_{n},\cdots, a_{1}\tau_{1})=\int W^{(n)}
(a_{n} a_{n}^{\prime} \tau_{n}, a_2 a_2^{\prime}\tau_2, \cdots
a_{1} a_{1}^{\prime} \tau_{1}) \,\,
da_1^{\prime}da_2^{\prime}\cdots da_{n}^{\prime}
\end{equation}
%-------------------------------------------------------------
and the correlation function is given by Eq.~(\ref{simplest2}).
Similarly, the response function is obtained by the following
integration
%-------------------------------------------------------------
\begin{equation}
\label{26a} R^{(n)} (\tau_{n} \cdots\tau_{1})=\int da_{1}\cdots
da_{n}\int da'_{1}\cdots da'_{n}(a_{n}-a_{n}^{\prime})\cdots
(a_{1}- a_{1}^{\prime}) W^{(n)} (a_{n} a_{n}^{\prime}
\tau_{n}\cdots a_{1} a_{1}^{\prime} \tau_{1}).
\end{equation}
%-------------------------------------------------------------

Comparing Eq.~(\ref{25}) with Eq.~(\ref{19}), it is clear that the
joint probability of measurements is related to the diagonal
elements of Eq.~(\ref{25}), i.e., $a_{j}=a'_{j}$. However we
cannot simply set $a_{j}=a'_{j}$ in Eq.~(\ref{25}) since it will
diverge. In order to properly obtain Eq.~(\ref{25}) from
Eq.~(\ref{19}) we need to add a finite resolution for the
measurement defined by a normalized function $f(a-a')$ sharply
peaked at zero. We can then write
%-------------------------------------------------------------
\begin{eqnarray}
\label{joint}& &  W_{S}^{(n)} (a_{n}\tau_{n}\cdots
a_{1}\tau_{1})=\\\nonumber & &  \int \cdots \int da_{1}^{\prime}
\cdots da_{n}^{\prime} W (a_{1}a'_{1} \tau_{1},a_{2}a'_{2}
\tau_{2}\cdots a_{n}a'_{n} \tau_{n})f (a_{1}-a'_{1})\cdots f
(a_{n}-a'_{n})
\end{eqnarray}
%-------------------------------------------------------------
Note that the definition of $W_{S}^{(n)}$ is more clean in
Liouville space (Eq.~(\ref{19})) but requires some care in Hilbert
space.

We can now better appreciate the fundamental differences between
these multitime various quantities. $W_{C}$ (and $C^{(n)})$ depend
only on $a_j$, where $a_j^\prime$ are integrated out. This follows
from Eq.~(\ref{12a}) which only contains ``left'' superoperators.
Because of the $a_j-a_j^\prime$ factors in Eq.~(\ref{26a})
$R^{(n)}$, on the other hand, depends only on the off diagonal
elements of $W$ with $a_j \neq a_j^\prime$ (diagonal elements do
not contribute). Finally, $W_S$ (and $S^{(n)})$ depends solely on
the diagonal elements of $W$ with $ a_j = a_j^\prime$. $W_S$ is
the basic quantity in the consistent history description of
quantum dynamics~\cite{gellmann,omnes1,omnes2} and has all the
properties of a classical joint probability distribution. At each
time $\tau_j$ the system is in the state $\mid\alpha_j>$ and its
density matrix is $\mid\alpha_j><\alpha_j\mid$ $j=0,\ldots n$. In
contrast, in a nonlinear response measurement as described by
$R^{(n)}$ we only measure $A$ at the last time $\tau_n$; at the
earlier times $\tau_j(j=1\ldots n-1)$ we only ``pass through''
$\alpha_j$, but the density matrix could be either
$\mid\alpha_j><\alpha_k\mid$ or $\mid\alpha_k><\alpha_j\mid$ with
$k\neq j$. $W_{R}$ is thus not a joint probability; even though we
perform some operation on the system at $n$ points ($n-1$
interactions with the fields and the time of observation), only
the last interaction corresponds to an actual measurement. In the
other times we merely perturb the system. It should be emphasized
that even though the response functions (and $W_{R}^{(n)}$) are
experimental observables that may be obtained from multiple pulse
experiments with heterodyne detection~\cite{shaul}, they may not
be represented by the joint probability $W_{S}^{(n)}$ since it
does not carry enough information for representing this type of
observables.

The response function carries information that depends on delicate
interferences among events that occur at the various points in
time. This interference may be understood in terms of sums over
pathways which differ by their time ordering i.e., $\langle
A(\tau_1) A(\tau_2) A(\tau_3)\rangle$, $\langle A (\tau_2)
A(\tau_1) A(\tau_3)\rangle$ etc. It is less obvious that a similar
interference does exist in classical mechanics as well.
Classically, of course, time ordering is immaterial since all
operators commute and it suffices to calculate $\langle A(\tau_1)
A(\tau_2) A(\tau_3)\rangle$ for $\tau_1 \le \tau_2 \cdots \le
\tau_n$. Quantum mechanically, each of the $n!$ permutations of
the time arguments in an $n$ point correlation function is
distinct and carries a different information. Classical
correlation functions then carry less information than their
quantum counterparts. The classical interference takes place
between closely lying trajectories~\cite{prl,300}.

\section{Correlation Function Expressions for Photon Statistics}

Photon counting statistics as well as shot noise statistics of
electrons~\cite{Beenakker} are most closely related to
$W_{S}^{(n)}$ (or $S^{(n)})$ since they involve $n$ real
measurements. However, photon counting is a more complex operation
than described by $W_{S}^{(n)}$ since it is performed under
nonequilibrium conditions where the system is strongly driven by
an external field. Furthermore, the material system is not closed
since photons are emitted. Thirdly, the measurement does change
the state of the system, not by merely projecting onto a diagonal
element. All of these complications can be adequately addressed
and Eq.~(\ref{19}) can be modified to account for photon
statistics, as will be shown below.

To proceed further we introduce the master equation, derived by
tracing the density matrix over the radiation
field~\cite{Lehmberg,AgarwalBook,43}. We shall denote by
$\Gamma_{\nu\nu'}$ the spontaneous emission rate from state $\nu'$
to the lower energy state $\nu$ . The total decay rate (inverse
radiative lifetime) of state $\nu'$ will then be
%-----------------------------------------------------------
\begin{equation}
\gamma_{\nu'}\equiv\sum_{\nu\neq\nu'}\Gamma_{\nu\nu'}.
\end{equation}
%-----------------------------------------------------------

The effects of spontaneous emission are then incorporated by the
master equation
%-----------------------------------------------------------
\begin{eqnarray}
\frac{d\rho_{\nu\nu'}}{dt}=-\frac{1}{2}(\gamma_{\nu}+\gamma_{\nu'})\rho_{\nu\nu'};
\hspace{0.2in}\nu\neq\nu'\\\nonumber
\frac{d\rho_{\nu\nu}}{dt}=-\gamma_{\nu} \rho_{\nu\nu} +
\sum_{\nu'\neq\nu}\Gamma_{\nu\nu'}\rho_{\nu'\nu'}.
\end{eqnarray}
%-----------------------------------------------------------
Adopting Liouville space (tetradic) notation, the master equation
reads
%-----------------------------------------------------------
\begin{eqnarray}
\label{33a} \frac{d\rho}{dt}&=&-\gamma\rho-\Gamma\rho,
\end{eqnarray}
%-----------------------------------------------------------
with
%-----------------------------------------------------------
\begin{eqnarray}
\label{344a}
\gamma&=&\sum_{\nu,\nu'}|\nu\nu'\gg\frac{1}{2}(\gamma_{\nu}+\gamma_{\nu'})\ll\nu\nu'|,
\end{eqnarray}
%-----------------------------------------------------------
and
%-----------------------------------------------------------
\begin{eqnarray}
\label{355a}
\Gamma&=&\sum_{\nu\neq\nu'}|\nu\nu\gg\Gamma_{\nu\nu'}\ll\nu'\nu'|.
\end{eqnarray}
%-----------------------------------------------------------
The quantities defined in the previous section correspond to a
closed-system and may be described either in Hilbert space or in
Liouville space~\cite{Molmer,Stenholm}; the choice is a matter of
convenience. The master equation allows us to avoid the explicit
treatment of the field degrees of freedom when describing an open
system. This can only be done for the density matrix in Liouville
space.

We consider a single quantum system coupled to a bath, driven by
an external field and subjected to spontaneous emission. The
Liouville equation will be partitioned as
%-----------------------------------------------------------
\begin{eqnarray}
\label{partitioned} \frac{d\rho}{dt}&=&-\frac{i}{\hbar} L(t)
\rho-\Gamma\rho.
\end{eqnarray}
%-----------------------------------------------------------
Here the Liouville operator $L(t)\equiv H_{-}(t) $ includes our
single multilevel system, any other bath degrees of freedom, as
well as the driving field. It also includes the $\gamma$ matrix
(Eq.~(\ref{344a})) which represents the diagonal (in Liouville
space) part of the master equation. $\Gamma$, on the other hand
(Eq.~(\ref{355a})), is the off diagonal part of the master
equation which describes the transitions among states $\nu$ and
$\nu'$.

We define the Green function solution of Eq.~(\ref{partitioned})
with $\Gamma=0$
%-----------------------------------------------------------
\begin{equation}
\label{337a} \mathcal{G} (\tau_{2},\tau_{1})\equiv T
\exp\left[-\frac{i}{\hbar}\int_{\tau_1}^{\tau_2}d\tau
L(\tau)\right],
\end{equation}
%-----------------------------------------------------------
and introduce the off diagonal radiative-decay operator in the
interaction picture
%----------------------------------------------------------------------
\begin{eqnarray}
\Gamma (\tau)\equiv \mathcal{G}^{\dag}(\tau,0)\Gamma
\mathcal{G}(\tau,0).
\end{eqnarray}
%-----------------------------------------------------------
The solution of Eq.~(\ref{partitioned}) in the interaction picture
then reads
%-------------------------------------------------------------
\begin{equation}
\label{interaction} \rho(t)= T
\exp\left[-\int_{t_{0}}^{t}d\tau\Gamma(\tau)\right]\rho(t_{0}).
\end{equation}
%-------------------------------------------------------------
The total probability density of emitting a photon between $t_{0}$
and $t_{0}+ dt_{0}$ and another photon between $t$ and $t+ dt$ is
%-------------------------------------------------------------
\begin{equation}
W(t,t_{0})=\mathrm{Tr}\Bigg[T\,\,\Gamma(t)\exp
\left[-\int_{t_{0}}^{t}d\tau\Gamma(\tau)\right]\Gamma(t_{0})\rho(t_{0})\Bigg].
\end{equation}
%-------------------------------------------------------------

Expanding the solution of Eq.~(\ref{interaction}) to $n$'th order
in $\Gamma$ yields
%-----------------------------------------------------------
\begin{equation}
\rho(t)=\sum_{n=0}^{\infty}\rho^{(n)}(t)
\end{equation}
%-----------------------------------------------------------
where
%-----------------------------------------------------------
\begin{eqnarray}
\rho^{(n-1)}(\tau_{n})
=\int_{0}^{\tau_{n}}d\tau_{n-1}\cdots\int_{0}^{\tau_{2}}d\tau_{1}\\\nonumber
\mathcal{G}(\tau_{n},\tau_{n-1})\Gamma\cdots\Gamma\mathcal{G}
(\tau_{2},\tau_{1})\Gamma\rho(\tau_{1}).
\end{eqnarray}
%-----------------------------------------------------------
$\rho^{(n-1)}$ describes $n-1$ photon emission processes at times
$\tau_{1}\cdots \tau_{n-1}$.

The probability density $K^{(n)}$ of emitting $n$ photons at times
$\tau_{1}\cdots \tau_{n}$ is obtained by multiplying the integrand
by $\Gamma$ from the left and taking a trace. This gives
%-----------------------------------------------------------
\begin{eqnarray}
\label{35a} K^{(n)}(\tau_{n}\cdots\tau_{1})=
\mathrm{Tr}\left[\Gamma (\tau_{n})\cdots \Gamma
(\tau_{1})\rho(\tau_{1})\right].
\end{eqnarray}
%-----------------------------------------------------------

This general expression for photon statistics in terms of system
variables is equivalent to the standard normally-ordered
expression of field
variables~\cite{Kelley,MandelBook,Glauber,Glauber2,Glauber3}. To
see that, we consider the probability of emitting $n$ photons
between times $t_{i}$ and $t_{f}$
%-----------------------------------------------------------
\begin{eqnarray}
\label{100aa} P_{n}(t_{f},t_{i}) =
\int_{t_{i}}^{t_{f}}d\tau_{n}\int_{t_{i}}^{\tau_{n}}d\tau_{n-1}\cdots
\int_{t_{i}}^{\tau_{2}}d\tau_{1}K^{(n)}(\tau_{n}\cdots\tau_{1})
\end{eqnarray}
%-----------------------------------------------------------

As can be seen from Eq.~(\ref{interaction}), this is normalized as
%-----------------------------------------------------------
\begin{eqnarray}
\label{101a} \sum_{n=0}^{\infty}P_{n}(t_{f},t_{i}) = \mathrm{Tr}
\rho(t)=1
\end{eqnarray}
%-----------------------------------------------------------
To compute $P_{n}$ we introduce the generating function
%-----------------------------------------------------------
\begin{eqnarray}
\label{47aa} G (t_{f},t_{i};U)\equiv
\sum_{n=0}^{\infty}P_{n}(t_{f},t_{i})U^{n}
\end{eqnarray}
%-----------------------------------------------------------
It then follows from Eqs.~(\ref{35a}) and Eq.~(\ref{100aa}) that
%-----------------------------------------------------------
\begin{eqnarray}
\label{48aa} G (t_{f},t_{i},U)=\left\langle \mathrm{T}\exp \left[U
\int_{t_{i}}^{t_{f}}d\tau \Gamma (\tau)\right]
\rho(t_{i})\right\rangle
\end{eqnarray}
%-----------------------------------------------------------
$G$ thus satisfies the equation of motion.
%-----------------------------------------------------------
\begin{eqnarray}
\frac{dG(t,t_{i};U)}{dt}=-\frac{i}{\hbar}L(t)G(t,t_{i},U)- U\Gamma
G(t,t_{i},U)
\end{eqnarray}
%-----------------------------------------------------------
Using Eq.~(\ref{48aa}) we have
%-----------------------------------------------------------
\begin{eqnarray}
\label{new99}
P_{n}(t,t+T)&=&\frac{1}{n!}\frac{d^{n}}{dU^{n}}G(U)\bigg|_{U=0}
\end{eqnarray}
%-----------------------------------------------------------
and Eq.~(\ref{47aa}) gives
%-----------------------------------------------------------
\begin{eqnarray}
\label{new100} \frac{d^{m}}{dU^{m}}G(U)\bigg|_{U=1}&=&
\sum_{n=0}^{\infty}P_{n}(t,t+T)\frac{n!}{(n-m)!}\\\nonumber
&\equiv&\langle n(n-1)\cdots(n-m+1)\rangle
\end{eqnarray}
%----------------------------------------------------------
which is the $n$'th factorial moment of $P_{n}$. Setting $m=1$ and
$m=2$ in Eq.~(\ref{new100}) we get
%----------------------------------------------------------
\begin{eqnarray}
\langle n \rangle =\frac{dG(U)}{dU}\bigg|_{U=1}\\\nonumber \langle
n^{2} \rangle-\langle n \rangle
=\frac{d^{2}G(U)}{dU_{2}}\bigg|_{U=1}
\end{eqnarray}
%----------------------------------------------------------
Higher moments may be calculated similarly.

The most commonly used measure of photon statistics, the Mandel
parameter, has been shown to be related to
$K^{(2)}$~\cite{Kelley,Leibfried,Metz} in simple kinetic models of
single quantum systems. Photon statistics has been calculated
using the simplest reduced descriptions such as the Bloch
equations~\cite{Zumofen}. The present approach opens up the use of
a broad class of reduced descriptions and stochastic
models~\cite{Kubo} for computing $K^{(2)}$. One interesting
application will be to photon statistics in superradiance of
aggregates~\cite{Science}. Eq.~(\ref{35a}) may be easily
generalized to describe more refined, frequency resolved,
measurements whereby at each time we monitor a different
(preselected) transition. This can be done simply by using a
different element of $\Gamma$ at each time $\Gamma(\tau_{j})$ to
represent the desired transition. Eq.~(\ref{35a}) could then
provide more detailed information about the system.

\section{Discussion}

We have introduced several types of multipoint functions commonly
used in experimental observations and their theoretical analysis.
Using the Liouville space superoperators notation, we can recast
these various quantities in a formally similar form that
facilitates their comparison. Eq.~(\ref{12a}) can be written as
%------------------------------------------------------------
\begin{eqnarray}
\label{recast} C^{(n)}(\tau_{n}\cdots\tau_{1})=
\mathrm{Tr}[A_{L}\mathcal{G}(\tau_{n}-\tau_{n-1})A_{L}\cdots
\mathcal{G}(\tau_{2}-\tau_{1})A_{L}\rho_{eq}]
\end{eqnarray}
%------------------------------------------------------------
where $\mathcal{G}(\tau)$ is given by Eq.~(\ref{100a}). The
nonlinear response function (Eq.~(\ref{3.5t})) can be similarly
recast in the form
%-------------------------------------------------------------
\begin{equation}\label{Eq. 3}
R^{(n)}(\tau_{n}\cdots \tau_{1}) = \mathrm{Tr}[ A_{+}
\mathcal{G}(\tau_{n}, - \tau_{n-1}) A_{-} \mathcal{G} \cdots A_{-}
\mathcal{G} (\tau_{2}, -\tau_{1}) A_{-} \rho_{eq}].
\end{equation}
%-------------------------------------------------------------
The joint distribution of successive measurements (Eq.~(\ref{19}))
is written as
%-------------------------------------------------------
\begin{equation}
\label{43a} W_{S}^{(n)}(a_{n}\tau_{n}\cdots a_{1}\tau_{1}) =
\mathrm{Tr} \left[\hat{P}({a_{n}})
\mathcal{G}(\tau_{n}-\tau_{n-1})\hat{P}({a_{n-1}})\cdots
\hat{P}({a_{2}})\mathcal{G}(\tau_{2}-
\tau_{1})\hat{P}({a_{1}})\rho_{eq}\right].
\end{equation}
%-------------------------------------------------------
The probability density of observing consecutive photons
(Eq.~(\ref{35a})) is
%-----------------------------------------------------------
\begin{eqnarray}
\label{34a} K^{(n)}(\tau_{n}\cdots\tau_{1})=
\mathrm{Tr}\left[\Gamma \mathcal{G}(\tau_{n},\tau_{n-1})\cdots
\Gamma \mathcal{G}(\tau_{2},\tau_{1})\Gamma\rho(\tau_{1})\right]
\end{eqnarray}
%-----------------------------------------------------------
where $\mathcal{G}(\tau,\tau')$ is given by Eq.~(\ref{337a}).
Finally, the probability density of measuring $n$ photons at times
$\tau_{1}\cdots\tau_{n}$ (regardless of how many photons are
emitted in between) is
%-------------------------------------------------------------
\begin{equation}
\label{46a}
P^{(n)}(\tau_{n}\cdots\tau_{1})=\mathrm{Tr}\left[\Gamma\tilde{\mathcal{G}}
(\tau_{n},\tau_{n-1})\cdots
\tilde{\mathcal{G}}(\tau_{2},\tau_{1})\Gamma\rho(\tau_{1})\right]
\end{equation}
%---------------------------------------------------------------
where $\tilde{\mathcal{G}}$ is the Green function solution of
Eq.~(\ref{partitioned}) for the driven system
%-------------------------------------------------------------
\begin{equation}
\rho(t) = \tilde{G} (t,t_{0})\rho(t_{0}).
\end{equation}
%--------------------------------------------------------------

We note several marked differences between the photon statistics
observables (Eqs.~(\ref{34a}) and ~(\ref{46a})) and the other
quantities (Eq.~(\ref{recast}),~(\ref{Eq. 3}) and (\ref{43a})).
Since the latter are equilibrium properties, the Green function is
translationally invariant and only depends on the time difference
$\mathcal{G}(\tau_{j}-\tau_{k})$ rather than on $\tau_{j}$ and
$\tau_{k}$ separately $\mathcal{G}(\tau_{j},\tau_{k})$. Also the
initial density matrix $\rho(\tau_{1})$ in photon statistics
measurements is generally a more complex object than $\rho_{eq}$
since it requires computing the preparation stage leading to a
nonequilibrium steady state. This does not cause any problem in
stochastic models where the bath evolution does not depend on the
state of the system. $\rho(\tau_{1})$ is then completely specified
since the first photon emission at $\tau_{1}$ determines the state
of the system (the final state of the emission) and the bath is
always in equilibrium. However, fully microscopic modelling will
require a separate calculation of $\rho(\tau_{1})$.

Eq.~(\ref{34a}) is very similar to the general expression for $n$
successive measurements (Eq.~(\ref{43a})). However, the $\Gamma$
matrix is off diagonal since photon emission is accompanied by a
transition in the system, as opposed to the diagonal
$\hat{P}({a})$ in Eq.~(\ref{43a}) which represents ordinary
measurements. Were we to use a diagonal
$\Gamma=|\nu\nu\gg\ll\nu\nu|$ it would represent the probability
of measuring the system at state $\nu$ at times $\tau_{1}\cdots
\tau_{n}$. Photon counting, however, implies that the system is at
state $\nu$ prior to the count but it changes to state $\nu'$
after the count; this is the initial state for the next period of
propagation. Apart from this, Eq.~(\ref{34a}) or Eq.~(\ref{46a})
are equivalent to $n$ point measurements (Eq.~(\ref{43a})). These
differences stem from the nonequilibrium nature of photon counting
performed on open driven systems.

Finally we note that Eq.~(\ref{34a}) and Eq.~(\ref{46a}) are
reminiscent of the normally ordered expressions with field
operators~\cite{Kelley, Glauber2} where $\Gamma$ represents the
detector rather than spontaneous emission. In the present approach
we do not need normal ordering since in Liouville space time
ordering is enough to maintain the bookkeeping of interactions. We
also note that $L(\tau)$ in Eq.~(\ref{337a}) contains the $\gamma$
matrix and the Green function therefore contains some diagonal
signatures of the photon emission. This is required for
maintaining the trace of the density matrix. Such terms should
also be present in the field formulation, but are usually
neglected and the Green function represents the pure system
(without the detector)~\cite{Kelley, Glauber2}. Adding these
corrections could improve the standard theory of photon
statistics.

\section{Acknowledgement}
The support of the National Science Foundation grant no.
(CHE-0132571) is gratefully acknowledged.

%-------------------------------------------------------------------------
%  REFERENCES
%-------------------------------------------------------------------------

\end{document}